\documentclass[11pt,a4paper,twoside]{article}  %
\usepackage[american]{babel}     
\usepackage{amsmath,amsthm,amssymb,latexsym}
\usepackage{amsfonts}           
\usepackage{geometry}
\usepackage{graphicx}         
\usepackage{fancyhdr}
\usepackage{tikz}
\usepackage{color}

\def\RR{\mathbb{R}}

\def\CC{\mathbb{C}}

\newtheorem{theorem}{Theorem}[section]    
\newtheorem{lemma}{Lemma}[section]   
\newtheorem{proposition}{Proposition}[section]
\newtheorem{corollary}{Corollary}[section]    
\newtheorem{remark}{Remark}[section]   
   
\newtheorem*{notation}{Notation}

\newenvironment { abstract }

\author{Domenico Felice\\
School of Science and Technology \\University of Camerino, I-62032 Camerino, Italy\\
INFN-Sezione di Perugia, Via A. Pascoli, I-06123 Perugia, Italy
\and 
Luigi Barletti\\
Dipartimento di Matematica e Informatica ``U.~Dini''\\   Universit\`a di Firenze, Italy}

\begin{document} 
\title{ \bf Correspondence between the NLS equation for optical fibers and a class of integrable NLS equations}

\maketitle

\begin{abstract}
The propagation of the optical field complex envelope in a single-mode fiber is governed by a one-dimensional cubic nonlinear 
Schr\"odinger equation with a loss term.
We present a result about $L^2$-closeness of the solutions of the above-mentioned equation and 
of a one-dimensional nonlinear Schr\"odinger equation that is Painlev\'e integrable.
\end{abstract}


\vspace{-6pt}

\section{Introduction}
The propagation of the optical field complex envelope $v(z,t),\ z,t\in \RR$ in a single-mode fiber, accounting for group velocity dispersion and Kerr nonlinearity, is governed by the nonlinear Schr\"odinger equation (NLSE) \cite{A}. 
In a time frame moving with the signal group velocity and for a normalized field $u(z,t):= e^{\alpha z/2}v(z,t)$(where $\alpha>0$ is the fiber loss coefficient), the NLSE can be written as
\begin{equation}
 i u_z = -\frac{\beta_2}{2} u_{tt} + \gamma e^{-\alpha z}|u|^2u,
\label{NLSE}
\end{equation}
where $\gamma>0$ is the Kerr coefficient, and $\beta_2\in\mathbb{R}\backslash\{0\}$ is the chromatic dispersion coefficient, 
usually given in square picosecond per kilometer \cite{A}. 
Note that, since $\beta_2$ can be either positive (normal dispersion) or negative (anomalous dispersion), the NLSE \eqref{NLSE}
can have a defocusing or focusing character, respectively.
We remark that with respect to the usual NLSE, here the space and time variable are exchanged and the longitudinal 
space variable $z$ is the evolution variable. 
This is the notation that will be adopted throughout this paper.
\par
Because of its dissipative character, Eq. \eqref{NLSE} is not integrable and, moreover, it does not admit solitonic solutions. 
However we shall show that a class of solutions of Eq. \eqref{NLSE} is close to solutions of an integrable NLSE. 
\par
\smallskip
In Ref.\ \cite{Weiss} is defined a Painlev\'e property for partial differential equations by extending the definition of the Painlev\'e property for ordinary differential equations; the latter is defined as follows. 
The solutions of a system of ordinary differential equations are regarded as analytic functions of a complex variable,
whose singularities depend on the initial conditions and are called, hence, ``movable''. 
The ordinary differential system is said to possess the Painlev\'e property when all the movable singularities are single-valued (simple poles).
\par
One major difference between analytic functions of one complex variable and analytic functions of several complex variables is that, in general, the singularities of a function of several complex variables cannot be isolated. 
If $f = f (\xi_1,\cdots,\xi_N )$ is a meromorphic function of $N$ complex variables ($2N$ real variables), the singularities of $f$ occur along analytic manifolds of (real) dimension $2N-2$. 
These manifolds are determined by conditions of the form $\phi(\xi_1,\cdots,\xi_N)=0$ where $\phi$ is an analytic function of
$(\xi_1,\cdots,\xi_N)$ in a neighborhood of the manifold.
\par
Therefore, we say that a partial differential equation has the Painlev\'e property when the solutions of the PDE are single-valued about the movable singularity manifold \cite{Weiss}. 
To be more precise, if the singularity manifold is determined by the condition $\phi(\xi_1,\ldots,\xi_n)=0,$ and $u(\xi_1,\ldots,\xi_n)$ 
is a solution of PDE, then we assume that
$$
u=u(\xi_1,\ldots,\xi_n)=\phi^q\sum_{j=0}^\infty u_j\phi^j,
$$
where $\phi$ and $u$ are analytic functions of $(\xi_1,\ldots, \xi_n)$ in a neighborhood of the singularity manifold, and $q$ is an integer.
Substitution of such expression of $u$ into the PDE determines the possible values of $q$ and determines a recursion relation for 
$ u_j,\; j=0,1,2,\ldots $ \cite{He}.  
This procedure determines solutions for PDEs that have the Painlev\'e property. 
This analysis is performed for general non autonomous NLSE with cubic nonlinearity \cite{He}.   
In Ref.\ \cite{Weiss} it is indicated that the Painlev\'e property may provide a unified description of integrable behavior in dynamical systems, 
while, at the same time, providing an efficient method for determining the integrability of particular systems. 
The integrability condition determines relations between the coefficients of the NLSE in order to have an integrable NLSE. 
It follows that it is possible to convert such integrable NLSE into the standard autonomous NLSE with cubic nonlinearity. 
The latter possesses analytic solutions; therefore it is also possible to obtain analytic solutions for a non-autonomous NLSE \cite{He}.
\par
The aim of this work is to identify a Painlev\'e-integrable NLSE which will be proven to be close to Eq.\ \eqref{NLSE} in $L^2$-norm. After this selection we can convert the last equation into the standard cubic NLSE. In order to obtain the result of closeness we make use of some arguments usually employed to prove finite-time blow up of solutions of the standard cubic NLSE \cite{C}. 
\par
The layout of this article is as follows. In Section \ref{section2} we briefly describe the Painlev\'e analysis that allows to obtain a family 
of integrable non autonomous NLSE; then we select the equation that will be proven to be close to Eq.\ \eqref{NLSE}.
Finally, we describe the transformation that allows us to pass to the standard cubic NLSE. In section \ref{section3} we state and prove the 
main result of correspondence. Finally, in section \ref{section4} we draw some conclusions by outlining the results obtained in this work 
and discussing possible extensions.
\section{The selection of the Painlev\'e integrable NLSE}
\label{section2}
To perform the analysis of the Painlev\'e test of integrability, that for partial differential equations is the well known 
Weiss-Tabor-Carnevale (WTC) test, we consider a fairly general form of the non-autonomous NLSE with real coefficients:
\begin{equation}
iv_z+fv_{tt}+g|v|^2v+Vv+ ihv=0.
\label{GNLSE}
\end{equation}  
Here $f(z,t)$ and $g(z,t)$ ($z,t$ real variables) are, respectively, the dispersion and the nonlinearity parameters to be determined. 
The function $V(z,t)$ is a potential-like term (which, in the context of optical fibers should be rather interpreted as due 
to a varying refraction index), and $h(z,t)$ is a dissipation ($h>0$) or gain ($h<0$) function. 
In Ref.\ \cite{He} the Authors find conditions which guarantees that Eq.\ \eqref{GNLSE} passes the WTC test. 
They employ the ansatz of Kruskal and take the singularity manifold in the form $\varphi(z,t):=t+\varphi(z)$. 
These conditions are summarized in the following proposition \cite{He}.
\begin{proposition}
Let $f(z,t),g(z,t),h(z,t),V(z,t)$ be analytic functions of $(z,t)$ in a neighborhood of the movable singularity manifold $\varphi(z,t):=t+\varphi(z)=0,$ then Eq. \eqref{GNLSE} is integrable in the sense of Painlev\'e   if $f,g,h$ are independent 
of $t$, and $\frac{\partial^3}{\partial t^3}V=0$, that is equivalent to require that
\begin{equation}
f(z,t)=f(z), \qquad g(z,t)=g(z),\qquad h(z,t)=h(z),\qquad V(z,t)=V_0(z)+V_1(z)t+V_2(z)t^2,
\label{compatibility1}
\end{equation}
where $V_0(z)$ and $V_1(z)$ are arbitrary. 
Moreover $f,g,h,V_2$ satisfy the following condition
\begin{equation}
(4f^2gg_z-2ff_zg^2)h-4f^2g^2h^2-2f^2g^2h_z-g^2ff_{zz}
+ f^2gg_{zz}-2f^2g_z^2 +f_z^2g^2+f_zgfg_z+4V_2f^3g^2=0.
\label{Relazione1}
\end{equation}
\end{proposition}
Conditions \eqref{compatibility1} imply that the Painlev\'e integrable class of Eq.\ \eqref{GNLSE} 
should have the form
\begin{equation} 
 iv_z(z,t)+f(z) v_{tt}(z,t)+g(z)|v(z,t)|^2v(z,t) + [V_0(z)+V_1(z)t+V_2(z)t^2]v(z,t)+ih(z)v(z,t)=0,
\label{GNLSE0}
\end{equation}
where $f(z),g(z),h(z)$ and $V_2(z)$ are related by \eqref{Relazione1}, and $V_0(z)$, $V_1(z)$ are arbitrary.
\begin{remark}
If we perform the transformation
$$
  u(z,t)=v(z,t)\exp\Big[-\int_0^z h(z^\prime)dz^\prime \Big],
$$
for every $z,t\in \RR$, Eq.\ \eqref{GNLSE0} becomes 
\begin{equation}
iv_z(z,t) +f(z) v_{tt}(z,t) + g(z)|v(z,t)|^2v(z,t)
+[V_0(z)+V_1(z)t+V_2(z)t^2]v(z,t)=0,
\label{GNLSE1}
\end{equation}
where $g(z)$ has been redefined accordingly,
and so the loss/gain term can be formally eliminated. 
Thus,  in what follows we only consider the model with $h(z)\equiv 0.$
In this case the relation between $V_2$, $f$ and $g$ becomes
\begin{equation}
-g^2ff_{zz}+f^2gg_{zz}-2f^2g_z^2+g^2f_z^2+gfg_zf_z+4V_2f^3g^2=0.
\label{relation2}
\end{equation}
\end{remark}
We are now ready to select the equation that we will show to be close to the NLSE for optical fibers, Eq. \eqref{NLSE}.
By choosing 
 $$
   f(z)=\frac{\beta_2}{2},\qquad g(z)=-\gamma e^{-\alpha z}
$$ 
in Eq.\ \eqref{GNLSE1} we obtain from \eqref{relation2} that 
$$
 V_2(z)=\frac{\alpha^2}{2\beta_2}.
$$
Moreover, without loss of generality we choose $V_0(z)\equiv V_1(z)\equiv 0,$ and, thus, we have 
\begin{equation}
iv_z(z,t)+\frac{\beta_2}{2}v_{tt}(z,t)-\gamma e^{-\alpha z}|v(z,t)|^2v(z,t)+\frac{\alpha^2}{2\beta_2}t^2v(z,t)=0
\label{NLSEintegrable}
\end{equation}
that is the Painlev\'e integrable NLSE that will be shown to be close to the NLSE \eqref{NLSE} in the $L^2$-distance.
\begin{remark}\label{remchoice}
Among all the possible equations \eqref{GNLSE1} we chose $f(z)=\frac{\beta_2}{2},\ g(z)=-\gamma e^{-\alpha z}$ in order to select an equation with the same dispersion and nonlinear effect of the Eq. \eqref{NLSE}. 
With this choice, it follows from Painlev\'e analysis that we have to consider also the additional potential term $V(z,t):=V_0(z)+V_1(z)t+V_2(z)t^2$, where $V_0$ and $V_1$ are arbitrary functions of the evolution variable $z$, while $V_2(z)$ has to satisfy relation \eqref{relation2}. 
It is clear that different choices for $f(z)$ and $g(z)$ lead to different expressions for the function $V_2(z)$; in general, from relation \eqref{relation2} it follows that there is a two-fold family of $C^{\infty}(\RR)$ functions 
yielding a Painlev\'e integrable NLSE. 
In the case under consideration, the selection $V_0(z)=0$ and $V_1(z)=0$ has been done for a sake of simplicity but
we cannot exclude that such choice is not optimal for the closeness estimations.
\end{remark}
\section{The correspondence theorem}
\label{section3}
Before stating the main result, we recall the physical dimensions of various quantities appearing in Eq.\ \eqref{NLSE} \cite{A}:
$$
  [\alpha]=L^{-1},\qquad[\beta_2]=L^{-1}T^2,\qquad [\gamma]=W^{-1}L^{-1},\qquad [u]=\sqrt{W},
$$
where $L$ is length, $T$ is time and $W$ is power.
By introducing a time scale normalized to the input width $T_0$ and a space scale  normalized to the fiber length $L_0,$ in the particular propagation regime where $L_0\sim L_D\sim L_{NL},$ where 
$$
L_D=\frac{T_0^2}{|\beta_2|},\qquad L_{NL}=\frac{1}{\gamma P_0}
$$
are the dispersion length $L_D$ and the nonlinear length $L_{NL},$ respectively, we bring Eq.\ \eqref{NLSE} into the equation
\begin{equation}
  iu_z(z,t)+ u_{tt}(z,t)+C_1e^{-C_2 z} |u(z,t)|^2u(z,t)=0
\label{modello}
\end{equation}
where $C_1=\pm 1$ (according to the focusing and defocusing character, respectively) 
and $C_2>0.$ 
Similarly, we bring Eq.\ \eqref{NLSEintegrable} into the equation
\begin{equation}
iv_z(z,t)+ v_{tt}(z,t)+C_1e^{-C_2 z} |v(z,t)|^2v(z,t)+\frac{C_2^2}{4}t^2v(z,t)=0
\label{modellointegrabile}
\end{equation}
(with the same $C_1$ and $C_2$).
The main result we are going to prove is the following.
\begin{theorem}
Let $v_0\in H^2(\RR)$ such that the functions $t^2v_0(t)$ and $tv'_0(t)$ belong to $L^2(\RR)$. 
Let $I_0 = [0,L_\mathrm{max})$ (with possibly $L_\mathrm{max} = +\infty$) be an interval of $\RR$ such that 
the problems
\begin{equation}
\left\{
\begin{aligned}
&iv_z+ v_{tt}+C_1e^{-C_2 z} |v|^2v+\frac{C_2^2}{4}t^2v=0
\\
&v(0,t)=v_0(t)
\end{aligned}
\right.
\label{cauchyint}
\end{equation}
and
\begin{equation}
\left\{\begin{aligned}
&iu_z+ u_{tt}+C_1e^{-C_2 z} |u|^2u=0
\\
&u(0,t)=v_0(t)
\end{aligned}\right.
\label{cauchying}
\end{equation}
have solutions $v \in L^\infty(I_0,H^2(\RR))$ and  $u \in L^\infty(I_0,H^2(\RR))$.
Then, for every $\varepsilon >0$ there exists $0< L(\varepsilon) \leq L_\mathrm{max}$ such that for every $0\leq L < L(\varepsilon)$ 
there exists $\delta >0$, depending on $\varepsilon$ and $L$ such that, if 
\begin{equation}
\label{initbound}
  ||t^2v_0||_{L^2}<\delta,\qquad ||t v'_0||_{L^2}<\delta,
\end{equation}
we have
\begin{equation}
  ||v(z)-u(z)||_{L^2}<\varepsilon
\end{equation}
for every $z\in [0,L].$
\label{maintheorem}
\end{theorem}
The proof of Theorem \eqref{maintheorem} is deferred to Subsection \ref{proof}.
In the proof, the $L^2$-distance between the two $H^2$-solutions will be evaluated in a natural way by 
considering the Duhamel formulas 
\begin{equation}
u(z) = \mathcal{T}(z)v_0
  +iC_1\int_0^z e^{-C_2 z^\prime}\mathcal{T}(z-z^\prime)\big[|u(z^\prime)|^2u(z^\prime)\big] dz^\prime,
\label{ingintegrale}
\end{equation}
and
\begin{equation}
v(z) = \mathcal{T}(z)v_0
    + iC_1\int_0^z e^{-C_2 z^\prime}\mathcal{T}(z-z^\prime)\big[|v(z^\prime)|^2v(z^\prime)\big]dz^\prime
  + i\frac{C_2^2}{4}\int_0^z\mathcal{T}(z-z^\prime)t^2v(z^\prime),
\label{intintegrale}
\end{equation}
where $\mathcal{T}(z) = \exp(\frac{z}{i}\frac{\partial^2}{\partial t^2})$ is given by
$$
\mathcal{T}(z)\varphi(t)=(4\pi i |z|)^{-1/2}\int_{\RR}e^{\frac{i|t-s|^2}{4z}}\varphi (s)ds.
$$
Here, with a little abuse of notation, we have denoted by $t^2v(z)$ the function $t \mapsto t^2v(z,t)$.
\par
It is clear that Eq.\ \eqref{intintegrale} is meaningful in the $L^2$ functional space if and only if the function 
$t\mapsto t^2v(z,t)$ belongs to $L^2(\RR)$ for every $z$  in a suitable interval $I$. 
Therefore, the first step is to find conditions so that
$$
  \int_{\RR}|t^2v(z,t)|^2 dt < \infty, \qquad\forall\ z\in I.
$$
\subsection{The correspondence with the standard cubic NLSE}
\label{corresp}
We now point out that, as it is well-known, the cubic NLSE is Painlev\'e integrable. 
So it is reasonable to look for the transformation that converts Eq.\ \eqref{modellointegrabile} into 
\begin{equation}
iQ_Z(Z,T)+Q_{TT}(Z,T)+\rho|Q(Z,T)|^2Q(Z,T)=0,
\label{modellocubica}
\end{equation}
whose solutions are global, that is they exist for every $Z\in[0,+\infty)$, both if $\rho$ is positive or negative  
\cite{C,Ka,Ts}.
In Ref.\ \cite{He} the right transformation is found in general for Eq.\ \eqref{GNLSE1}. 
In our specific case, this reads as follows
\begin{equation}
\left\{\begin{array}{l}
T(z,t)=e^{-C_2 z}t\\
\\
Z(z)=\frac{1}{2 C_2}(1-e^{-2C_2 z})\\
\\
v(z,t)=\;e^{i\frac{C_2}{4}t^2-\frac{C_2}{2}z}Q(Z(z),T(z,t))\\
\end{array}\right.
\label{trasformazionimodello}
\end{equation}
where $v(z,t)$ and $Q(Z,T)$ solve Eq.\ \eqref{modellointegrabile} and Eq.\ \eqref{modellocubica}, respectively, 
and $z\in I$, $t\in \RR$, $C_1=\pm 1$, $C_2>0$.
\begin{remark}
It is easy to prove that transformation \eqref{trasformazionimodello} possesses the inverse transformation, that is
\begin{equation}
\left\{\begin{array}{l}
t(Z,T)=\Big(1-2C_2Z\Big)^{-1/2}T\\
\\
z(Z)=-\frac{1}{2C_2}\ln\Big(1-2C_2Z\Big)\\
\\
Q(Z,T)=\;e^{i\frac{C_2}{4}t^2-\frac{C_2}{2}z}v(z(Z),t(Z,T))\\
\end{array}\right.
\label{trasformazioniinversemodello}
\end{equation}
where $Z\in [0,\frac{1}{2C_2})$ and $T$ runs over $\RR.$   
\end{remark}
\begin{notation}
\label{intervallo}
As introduced in Theorem \ref{maintheorem}, hereafter we refer to $I_0$ as the interval $[0,L_\mathrm{max})$ where the solutions 
of the problems \eqref{cauchyint} and \eqref{cauchying} exist.
Under the transformations \eqref{trasformazionimodello}, $I_0$ becomes the interval
$$
   \widetilde{I} :=
   Z(I_0) = \big[0,\frac{1}{2 C_2}(1-e^{-2C_2 L_\mathrm{max}}) \big].
$$
\end{notation}
\begin{lemma}
Let  $v(z,t)$ be an arbitrary  solution of equation \eqref{modellointegrabile} on $I_0$, and $Q(Z,T)$ an arbitrary solution 
of equation \eqref{modellocubica} on $\widetilde{I}.$ 
Then the function $t \rightarrow t^2 v(z,t)$ belongs to $ L^2(\RR)$ for every $z\in I_0$ if and only if the function 
$T\mapsto T^2Q(Z,T)$  belongs to $ L^2(\RR)$ for every $Z\in \widetilde{I}$.
\label{shift}
\end{lemma}
\noindent
{\bf Proof}\ Thanks to transformations \eqref{trasformazioniinversemodello} it follows that
\begin{eqnarray*}
||t^2v(z)||_{L^2}^2&=&\int_{\RR}|t^2v(z,t)|^2 dt = e^{4C_2 z}\int_{\RR}|T^2Q(Z,T)|^2 dT.
\end{eqnarray*}
Therefore $||(\cdot)^2v(z,\cdot)||_{L^2}=e^{2C_2 z}||(\cdot)^2Q(Z(z),\cdot)||_{L^2}$, and the thesis follows. \hfill $\Box$
\subsection{A class of solutions of the integrable NLSE}
\label{classolution}
In what follows we carry out some arguments of Ref.\ \cite{C} in order to show that the function $T\mapsto T^2 Q(Z,T)$ belongs to 
$L^2(\RR)$, for every $Z$ in a suitable interval. 
Recalling that $Q(Z,T)$ is a solution of the NLSE \eqref{modellocubica}, we start by multiplying formally Eq.\ \eqref{modellocubica} 
by $|T|^4\overline{Q},$ where $\overline{Q}$ is the complex conjugate of $Q$:
\begin{equation}
iT^4Q_Z\overline{Q}+ T^4Q_{TT}\overline{Q}+\rho T^4|Q|^4=0,
\label{SNLSEmoltiplicata}
\end{equation}
where $\rho=\pm 1.$
We integrate over $\RR$ with respect to the variable $T$, then after an integration by parts and, finally, 
by taking the imaginary part we obtain
$$
\frac 1 2 \frac{d}{dZ}\int_{\RR}|T^2 Q|^2 dT=4\mbox{Im}\int_{\RR}(T^2\overline{Q})(TQ_T) dT.
$$
In the following Proposition we are going to prove the boundedness of the integral $\int_{\RR}T^2|Q_T(Z,T)|^2 dT$, 
for every $Z$ in a suitable interval, in order to be allowed to apply the Cauchy-Schwartz inequality to the right-hand side 
of the last equality. 
\begin{proposition}
Let $Q(Z,T)$ be a $H^2$  solution on a bounded interval $I = [0,L]$ of the standard NLSE \eqref{modellocubica}. 
If $T Q_T(0,T) \in L^2(\RR)$, then the function $Z \mapsto TQ_T(Z,T)$ belongs to $L^\infty(I,L^2(\RR))$.
 \label{proposizione}
\end{proposition}
\noindent 
{\bf Proof}\ Let us set $W(Z,T) := Q_T(Z,T)$.
By differentiating the standard NLSE \eqref{modellocubica} with respect to the variable $T$ we obtain:
$$
iQ_{ZT}+ Q_{TTT}+\rho (2|Q|^2 Q_T+Q^2\overline{Q}_T)=0,
$$
that is
\begin{equation}
iW_Z+ W_{TT}+\rho (2|Q|^2W+Q^2\overline{W})=0.
\label{W}
\end{equation}
Let us multiply formally Eq.\  \eqref{W} by $T^2\overline{W}$:
$$
iT^2W_Z\overline{W}+ T^2W_{TT}\overline{W}+\rho (2T^2|Q|^2|W|^2+T^2Q^2\overline{W}^2)=0.
$$
Integrating over $\RR $ with respect to the variable $T$, and taking the imaginary part, we have that
$$
\frac{1}{2} \frac{d}{dZ}\int_{\RR}T^2|W(Z,T)|^2 dT=
-\mathrm{Im} \left(\int_{\RR}T^2W_{TT}(Z,T)\overline{W}(Z,T)dT
+\rho \int_{\RR}T^2Q^2(Z,T) \overline{W}^2(Z,T) dT\right).
$$
Performing another integration by parts with respect the variable $T$ in the first integral on the right-hand side 
of the previous equation we arrive at
$$
\frac{1}{2} \frac{d}{dZ}\int_{\RR}T^2|W(Z,T)|^2 dT=
 \mathrm{Im}  \int_{\RR}T\overline{W}(Z,T)\Big[2 W_T(Z,T)-\rho T\overline{W}(Z,T)Q^2(Z,T)\Big] dT .
$$
Let us set $h(Z)=||TW(Z)||_{L^2}^2$ and rewrite the previous relation as
$$
h^\prime(Z)=2\mbox{Im}  \int_{\RR}T\overline{W}(Z,T)\Big[2 W_T(Z,T)-\rho T\overline{W}(Z,T)Q^2(Z,T)\Big] dT.
$$
We observe that this equation is not, strictly speaking, rigorous since it has been obtained by a formal calculation.
However it can be made rigorous straightforwardly by means of a suitable regularization procedure.
\par
By the fundamental theorem of calculus we have
$$
h(Z)=h(0)+\int_0^Z 2\mathrm{Im} \Big( \int_{\RR}T\overline{W}(Z^{\prime})\Big[2 W_T(Z^{\prime})
  -\rho T\overline{W}(Z^{\prime})Q^2(Z^{\prime})\Big] dT\Big) dZ^{\prime} .
$$
From the Cauchy-Schwartz inequality and the triangular inequality we have that
$$
h(Z)\leq h(0)+2 \int_0^Z ||TW(Z^{\prime})||_{L^2}\Big[||2 W_{T}(Z^{\prime})||_{L^2}
+||\rho T\overline{W}(Z^{\prime})Q^2(Z^{\prime})||_{L^2}\Big]dZ^\prime .
$$
Now by using H\"older inequality and the Sobolev immersion $H^2(\RR)\hookrightarrow L^\infty(\RR)$ we notice that
\begin{equation*}
||\rho T\overline{W}Q^2||_{L^2}\leq ||Q||_{L^\infty}^2||TW||_{L^2}.
\end{equation*}
So we obtain the relation
$$
h(Z)\leq h(0)+2\int_0^Z\Big[2||TW(Z^{\prime})||_{L^2}|| W_{T}(Z^{\prime})||_{L^2}+||TW(Z^{\prime})||_{L^2}^2||Q(Z^{\prime})||^2_{L^\infty}\Big]dZ^\prime.
$$
\par 
By assumption we have that $Q \in C(I,H^2(\RR)).$  
Since  $W=Q_T$, we have that $||W_T(Z)||_{L^2}$ is finite for every $Z\in I$.  
Moreover, since $||Q(Z)||_{L^\infty}\leq C||Q(Z)||_{H^2},$ with $C$ positive constant independent of $Z,$ we have that $||Q(Z)||^2_{L^\infty}$ 
is finite for every $Z\in I$. 
In this way we obtain that
\begin{equation}
h(Z)\leq h(0)+\widetilde{C}\int_0^Z\Big[\sqrt{h(Z^\prime)}+h(Z^\prime)\Big]dZ^\prime,
\label{gronwallh}
\end{equation}
where  $\widetilde{C}$ is independent of $Z$  ($Q(Z)$ could be extended over all $Z>0$). 
Relation \eqref{gronwallh} can be made explicit with respect the function $h(Z)$ by using a Gronwall argument.
Let us set  
$$
  u(Z) := h(0)+\widetilde{C}\int_0^Z\Big[\sqrt{h(Z^\prime)}+h(Z^\prime)\Big]dZ^\prime.
$$
From \eqref{gronwallh} it follows that $h(Z)\leq u(Z)$ for every $Z\in I$.  
Thus, we can write 
\begin{equation*}
u^\prime(Z) = \widetilde{C}\left[\sqrt{h(Z)}+h(Z)\right] \leq \widetilde{C}\left[\sqrt{u(Z)}+u(Z)\right]
\end{equation*}
Therefore by differentiating the function $\ln[\sqrt{u(Z)}+1]$ we obtain that
$$
\frac{d}{dZ}\ln[\sqrt{u(Z)}+1] = \frac{u^\prime(Z)}{2[\sqrt{u(Z)}+u(Z)]}
\leq \frac{\widetilde{C}}{2}.
$$
We now integrate from $0$ and $Z$, and we obtain that
$$
\ln[\sqrt{u(Z)}+1] \leq \ln[\sqrt{u(0)}+1]+ \frac{\widetilde{C} Z}{2},
$$
for every $Z \in I$.
Then, it follows that
$$
\sqrt{u(Z)}+1 \leq [\sqrt{u(0)}+1]\,e^{\frac{\widetilde{C} Z}{2}},
$$
and so, being $u(0)=h(0),$ we obtain that
\begin{equation}
  \sqrt{h(Z)} \leq [\sqrt{h(0)}+1]\,e^{\frac{\widetilde{C} Z}{2}}-1
\label{hGronwall}
\end{equation}
for all $Z\in I$.
Thus, we have obtained that the function $T\mapsto TW(Z,T)$ belongs to $L^2(\RR)$ for every $Z\in I$.
\hfill $\Box$
\bigskip
\par
The next result states the conditions under which the function $T^2Q(Z,T)$ belongs to $L^2(\RR)$ 
for every $Z\in I$.
\begin{proposition}
Let $Q_0(T)\in H^2(\RR)$ and $Q(Z,T)$ be a $H^2$-solution of
$$
\left\{
\begin{aligned}
&iQ_Z+ Q_{TT}+\rho|Q|^2Q= 0
\\
&Q(0,T)=Q_0(T)
\end{aligned}
\right.
$$
on a bounded interval $I = [0,L]$. 
If $T^2Q(0,T)\in L^2(\RR)$ and $TQ_T(0,T)\in L^2(\RR)$, then the function $Z\mapsto T^2 Q(Z,T)$ belongs to $L^\infty(I,L^2(\RR))$.
\label{teoremadishift}
\end{proposition}
{\bf Proof.}\ By multiplying formally the standard NLSE by $T^4\overline{Q}$, integrating over $\RR $ and 
taking the imaginary part we have that
$$
\frac 1 2 \frac{d}{dZ}\int_{\RR}|T^2 Q(Z,T)|^2 dT=4\mbox{Im}\int_{\RR}T^2\overline{Q}(Z,T)TQ_T(Z,T) dT
$$
Now, let us define $ f(Z)=||T^2Q(Z,\cdot)||_{L^2}^2;$ then  from the fundamental theorem of calculus we have that
$$
f(Z) =  f(0)+\int_0^Z 8\mbox{Im}\int_{\RR}T^2\overline{Q}(Z^\prime,T)TQ_T(Z^\prime,T) dT dZ^\prime
$$
Since  $TQ_T(0,T)\in L^2(\RR)$ by assumptions, it follows from Proposition \ref{proposizione} that the function $Z\mapsto TQ_T(Z,T)$ belongs to $L^\infty(I,L^2(\RR))$.  
From the Cauchy-Schwartz inequality it follows that 
\begin{equation}
f(Z)\leq f(0)+8\int_0^Z||TQ_T(Z^\prime)||_{L^2} \sqrt{f(Z^\prime)} dZ^\prime .
\label{stimaf}
\end{equation} 
Relation $(\ref{stimaf})$ can be made explicit with respect to $f(Z)$: let us set 
\begin{equation*}
u(Z) := f(0)+8\int_0^Z ||TQ_T(Z^\prime)||_{L^2} \sqrt{f(Z^\prime)}\, dZ^\prime .
\end{equation*}
By differentiating with respect to $Z$ we have that
$$
u^\prime(Z) =  8||TQ_T(Z)||_{L^2} \sqrt{f(Z)}
\leq 8||TQ_T(Z)||_{L^2}\, \sqrt{u(Z)},
$$
where we used the inequality \eqref{stimaf}.  
 Therefore by differentiating the function $\sqrt{u(Z)}$ we obtain that
\begin{equation*}
\frac{d}{dZ}\sqrt{u(Z)}=\frac{u^\prime(Z)}{2\sqrt{u(Z)}} \leq 4||TQ_T(Z)||_{L^2}.
\end{equation*}
We now integrate from $0$ to $Z$ and obtain
$$
\sqrt{u(Z)} \leq \sqrt{u(0)}+4\int_0^Z||TQ_T(Z^\prime)||_{L^2}\,dZ^\prime.
$$
We now observe that $u(0)=f(0)$ and, from $(\ref{stimaf})$ and the definition of $u(Z),$ it follows that  
$\sqrt{f(Z)}\leq \sqrt{u(Z)}$ for every $Z\in I.$ 
So, we conclude that
\begin{equation}
\sqrt{f(Z)}\leq \sqrt{f(0)}+4\int_0^Z||TQ_T(Z^\prime)||_{L^2}\,dZ^\prime,
\label{gronwallf}
\end{equation}
which proves the thesis. 
\hfill $\Box$
\par \bigskip
\begin{corollary}
Let  $v(z,t)$ as in Eq.\ \eqref{trasformazionimodello}, with $Q(Z,T)\in C(\widetilde{I},H^2(\RR))$ a solution of the
 standard NLSE \eqref{modellocubica}.
 If $t^2v(0,t)\in L^2(\RR)$ and $tv_t(0,t)\in L^2(\RR)$ then the function $z\mapsto t^2v(z,t)$ belongs to 
 $L^2(\RR)$ for every $z\in I$.
\label{corollary}
\end{corollary}
{\bf Proof.} 
In order to apply Proposition \ref{teoremadishift} we need $T^2Q(0,T)\in L^2(\RR)$ and 
$TQ_T(0,T)\in L^2(\RR)$. 
Making use of \eqref{trasformazioniinversemodello} we obtain that
$$
Q(0,T)=e^{-i\frac{C_2}{4}t^2(0,T)}v(0,t(0,T)).
$$
We differentiate this relation  with respect to $T$ and obtain
$$
Q_T(0,T) = e^{-i\frac{C_2}{4}t^2}\Big[-i\frac{C_2}{2}\,tv(0,t)+v_t(0,t)\Big],
$$
where we used  $t(0,T)=T$ and  $\frac{dt}{dT}=1$
By using the relation $T(0,t)=t$ we obtain that
$$
TQ_T(0,T)=t\,e^{-i\frac{C_2}{4}t^2}\Big[-i\frac{C_2}{2}tv(0,t)+v_t(0,t)\Big].
$$
We now integrate this relation over $\RR$ with respect the variable $T$ and, using $dT=dt$, we obtain that
\begin{eqnarray*}
\int_{\RR}|TQ_T(0,T)|^2 dT &\leq& \Big[\Big(\frac{C_2}{2}\Big)^2\int_{\RR}\Big|t^2v(0,t)\Big|^2 dt
+\int_{\RR}\Big|tv_t(0,t)\Big|^2 dt\Big],
\end{eqnarray*}
so from the assumptions we have that $TQ_T(0,T)$ belongs to $L^2(\RR).$
\par
Furthermore, we have that
\begin{eqnarray*}
\int_{\RR}\Big|T^2 Q(0,T)\Big|^2 dT &=&\int_{\RR}|t^2v(0,t)|^2 dt,
\end{eqnarray*}
so also $T^2 Q(0,T)$ belongs to $L^2(\RR)$. 
Therefore, by Proposition \ref{teoremadishift} we have that the function $Z\mapsto T^2Q(Z,T)$ belongs to 
$L^2(\RR)$ for every $Z\in I$. 
Finally, as pointed out in Lemma \ref{shift} we obtain that
$$
||t^2v(z)||_{L^2}= e^{2C_2 z}||T^2Q(Z)||_{L^2}
$$
and, therefore, the function $t\mapsto t^2v(z,t)$ belongs to $L^2(\RR)$ for every $z\in I$.
\hfill $\Box$
\par \medskip
We are now in position to prove the anticipated Theorem \ref{maintheorem}. 
\subsection{Proof of Theorem \ref{maintheorem}}
\label{proof}
By assumption, the functions $v(z,t)$ and $u(z,t)$ are solutions of the problems \eqref{cauchyint} and \eqref{cauchying},
respectively, and belong to $L^\infty(I_0,H^2(\RR))$ (with possibly $I_0 =  [0,+\infty)$ \cite{C,F}). 
Let us consider $f\in C^1(\CC,\CC)$ defined by $f(\zeta)=|\zeta|^2\zeta$. 
Then we have $|f(\zeta)-f(\xi)|\leq L(K)|\zeta-\xi|,$ for all $\zeta,\xi\in\CC$ such that $|\zeta|^2+|\xi|^2 \leq K$, with 
$L(t)\in C([0,\infty)$. 
Since $u$ and $v$ belong to $L^\infty(I_0,H^2(\RR))$, then $K>0$ exists such that $|u(z,t)|^2 + |v(z,t)|^2 \leq K$ for (almost)
all $(z,t) \in I_0 \times \RR$. 
This implies that a constant  $C>0$ exists such that $||f(u(z))-f(v(z))||_{L^2}\leq C||v(z,\cdot)-u(z)||_{L^2}$ for all  $z \in I_0$
(see also \cite{C,Ka}).
Recalling that the operator $\mathcal{T}(z) = \exp(\frac{z}{i}\frac{\partial^2}{\partial t^2})$ acts as an isometry on the space $L^2(\RR)$, using Corollary \ref{corollary} and the assumptions of Theorem \ref{maintheorem}, from the Duhamel formulas \eqref{ingintegrale} and \eqref{intintegrale} we obtain
\begin{equation}
||v(z)-u(z)||_{L^2}\leq C \int_0^z||v(z^\prime)-u(z^\prime)||_{L^2}\,dz^\prime 
+\frac{C_2^2}{4}\int_0^z||t^2v(z^\prime)||_{L^2}\, dz^\prime,
\label{duhamel}
\end{equation}
where $C$ and $C_2$ are positive constants, and we used the fact that $C_1 = \pm1$. 
The idea of the proof is to get an estimate of the distance between the functions $v$ and $u$ by using \eqref{duhamel}. 
First of all, we need to treat the last term at the right-hand side of Eq.\ \eqref{duhamel}. 
We know that
\begin{equation}
v(z,t)=e^{i\frac{C_2}{4}t^2-\frac{C_2}{2}z}Q(Z(z),T(z,t)),
\label{trans}
\end{equation}
where $Q(Z,T)$ is a solution of the standard NLSE \eqref{modellocubica}  for every 
$Z\in \widetilde{I} := Z(I_0)$.
Let us set $g(z):=||t^2v(z)||_{L^2}^2$ and $f(Z):=||T^2Q(Z)||_{L^2}^2.$ 
The relation between these functions was established into the proof of the Lemma $\ref{shift},$ that is
\begin{equation}
g(z)=e^{4C_2z}f(Z).
\label{norma}
\end{equation}
In the proof of Proposition \ref{teoremadishift} we have found that $f(Z)$ satisfies inequality \eqref{gronwallf}.
Moreover, in the proof of of Proposition  \ref{proposizione} we have shown that  $h(Z):= ||TQ_T(Z)||_{L^2}^2$
satisfies inequality \eqref{hGronwall}, where the constant $\widetilde{C} > 0$ is independent of $Z$.
  From the inverse transformation $(\ref{trasformazioniinversemodello})$ we can write
$$
Q(Z,T) = e^{-i\frac{C_2}{4}t^2(Z,T)+\frac{C_2}{2}z(Z)}v(z(Z),t(Z,T))
$$
and so
$$
Q_T(0,T) = e^{-i\frac{C_2}{4}t^2}\big[-i\frac{C_2}{2}tv(0,t)+v_t(0,t)\big].
$$
Therefore, assuming $||t^2v(0,t)||_{L^2} < \delta$ and $||tv_t(0,t)||_{L^2} < \delta$, we obtain
\begin{equation*}
||TQ_T(0,T)||_{L^2}\leq\frac{C_2}{2}||t^2v(0,t)||_{L^2}+||tv_t(0,t)||_{L^2}<\Big(\frac{C_2}{2}+1\Big)\delta.
\end{equation*}
Finally, by substituting the latter relation into \eqref{hGronwall} we have
\begin{equation}
||TQ_T(Z)||_{L^2} = \sqrt{h(Z)}<\Big[\Big(\frac{C_2}{2}+1\Big)\delta+1\Big]e^{\frac{\widetilde{C}}{2}Z}-1,
\label{stimah}
\end{equation}
where we notice that the right-hand side of the inequality is positive for every $Z\in \widetilde{I}$.
\par
Now, since $\sqrt{f(0)}=||t^2v(0)||_{L^2}$, we have that $\sqrt{f(0)}<\delta$, and from \eqref{gronwallf} we obtain
\begin{multline}
||T^2Q(Z)||_{L^2} = \sqrt{f(Z)} \leq \sqrt{f(0)}+4\int_0^Z\sqrt{h(Z^\prime)}dZ^\prime
\\
< \delta+4\int_0^Z\Big[\Big[\Big(\frac{C_2}{2}+1\Big)\delta+1\Big]e^{\frac{\widetilde{C}}{2}Z^\prime}-1 \Big]dZ^\prime
= \delta+\frac{8}{\widetilde{C}}\Big[\Big(\frac{C_2}{2}+1\Big)\delta+1\Big]\Big(e^{\frac{\widetilde{C}}{2}Z}-1\Big)-4 Z.
\label{Stima}
\end{multline}
Finally, using \eqref{norma} we can write
$$
\sqrt{g(z)} < e^{2C_2z}\Bigl\{\delta+\frac{8}{\widetilde{C}}\Big[\Big(\frac{C_2}{2}+1\Big)\delta+1\Big]
\Big(e^{\frac{\widetilde{C}}{2}Z}-1\Big)-4Z\Bigr\}
$$
and so
\begin{equation}
||t^2v(z)||_{L^2} = \sqrt{g(z)}<e^{2C_2z} \eta(Z(z),\delta),
\label{stimag}
\end{equation}
where 
\begin{equation}
\eta(Z,\delta)=\delta+\frac{8}{\widetilde{C}}\Big[\Big(\frac{C_2}{2}+1\Big)\delta+1\Big]\Big(e^{\frac{\widetilde{C}}{2}Z}-1\Big)-4Z,
\label{eta}
\end{equation}
where it is readily seen that $\eta(Z,\delta) > 0$ is an increasing function of $Z$. 
Going back to \eqref{duhamel}, we can write, therefore, 
\begin{equation}
||v(z)-u(z)||_{L^2} < C \int_0^z||v(z^\prime)-u(z^\prime)||_{L^2}dz^\prime 
 +\frac{C_2}{4}\eta(Z(z),\delta)z\,e^{2C_2z}.
\end{equation}
Recalling that $Z(z)$ is increasing on $I_0$, the function $\frac{C_2}{4}\eta(Z(z),\delta)z\,e^{2C_2z}$
is also increasing on $I_0$ and, then, we can apply the Gronwall Lemma and obtain that
\begin{equation}
||v(z)-u(z)||_{L^2}<\frac{C_2}{4}\eta(Z(z),\delta)z\,e^{(2C_2+C)z},
\label{Stimadistanza}
\end{equation}
for all $z \in I_0$.
Given $\varepsilon>0,$ we have to prove that there exists $\delta>0$ such that
\begin{equation*}
\frac{C_2}{4}\eta(Z(z),\delta)z\,e^{(2C_2+C)z} <\varepsilon.
\end{equation*}
Recalling \eqref{eta}, we need that $\delta$ fulfils
\begin{equation}
\label{delta}
 \left[1+\frac{8}{\widetilde{C}}\Big(\frac{C_2}{2}+1\Big)\Big(e^{\frac{\widetilde{C}}{2}Z(z)}-1\Big)\right]\delta
  <G(z,\varepsilon)-H(z),
\end{equation}
uniformly with respect to $z$ in a suitable interval, where
$$
G(z,\varepsilon):=\frac{4\,\varepsilon\, e^{-(2C_2+C)z}}{C_2\, z},
\qquad
H(z):=  \frac{8}{\widetilde{C}}\left(e^{\frac{\widetilde{C}}{2}Z(z)} -1\right)-4Z(z).
$$
Recalling that $Z(z) = \frac{1}{2C_2}\left(1 - e^{-2C_2z} \right)$, we have
$$
\begin{aligned}
&\lim_{z \to 0} G(z,\varepsilon) = +\infty, &\quad &\lim_{z \to +\infty} G(z,\varepsilon) = 0,
\\[4pt]
&\lim_{z \to 0} H(z) = 0, &\quad &\lim_{z \to +\infty} H(z) =  \frac{8}{\widetilde{C}}\Big(e^{\frac{\widetilde{C}}{4C_2}} -1\Big)
-\frac{2}{C_2},
\end{aligned}
$$
with $G(z,\varepsilon)$ monotonically decreasing and $H(z)$ monotonically increasing.
Thus, there exists $0< L(\varepsilon) \leq L_\mathrm{max}$ such that $G(z,\varepsilon)-H(z)$ is positive for every 
$0< z < L(\varepsilon)$.
Then, from \eqref{Stimadistanza} and \eqref{delta} we see that for any given $0 \leq L < L(\varepsilon)$, we have 
\begin{equation*}
||v(z)-u(z)||_{L^2} < \frac{C_2}{4}\eta(Z(L),\delta)L\,e^{(2C_2+C)L} < \varepsilon
\end{equation*}
for all $z \in [0,L]$, whenever
\begin{equation}
\label{deltacond}
 \delta < \left[1+\frac{8}{\widetilde{C}}\Big(\frac{C_2}{2}+1\Big)\Big(e^{\frac{\widetilde{C}}{2}Z(L)}-1\Big)\right]^{-1}
  \Big(G(L,\varepsilon)-H(L)\Big),
\end{equation}
which proves our thesis.
\hfill $\Box$
\par
\begin{remark}
\label{finalrem}
The proof of the Theorem \ref{maintheorem} basically provides a method to find approximate analytic solutions to the dimensionless fiber-optic NLSE \eqref{modello}.
We have proven that, given the error bound $\varepsilon$, a propagation distance $L(\varepsilon) > 0$ and a 
$\delta(\varepsilon) > 0$ exist such that, if the the initial datum is controlled by $\delta$ according to \eqref{initbound}, 
then the error is less than $\varepsilon$ within that propagation distance. 
However, the proof of Theorem \ref{maintheorem} can be used in the reverse way, i.e.\ for estimating the error for
a given propagation distance $\bar{L}$: if $\varepsilon$ is such that $G(\bar{L},\varepsilon) > H(\bar{L})$, then,
provided that the initial datum is controlled by the $\delta$ given by \eqref{deltacond} with $L=\bar{L}$, the error will be 
at most $\epsilon$.
\end{remark}
\section{Conclusions}
\label{section4}
The main result obtained in this work is Theorem \ref{maintheorem} which states that the dissipative NLSE of optical fibers,
which is not integrable, is close in $L^2$ norm to a Painlev\'e-integrable NLSE. The Theorem \ref{maintheorem} basically provides a method to find approximate analytic solutions to the fiber-optic NLSE \eqref{NLSE}, bounding the error for a given propagation distance, as highlighted in Remark \ref{finalrem}. 
The main result has been proven for the dimensionless version \eqref{modello} of Eq.\ \eqref{NLSE}. 
As discussed in Remark \ref{remchoice}, the Painlev\'e condition leaves us with a certain amount of freedom in the choice of the approximating integrable NLSE, and our choice of Eq.\ \eqref{NLSEintegrable} has been dictated by intuitive considerations. 
Such choice works well but we do not claim that it is optimal. 
Also the estimations of Theorem  \ref{maintheorem} are not intended to provide optimal expressions and constants.
\par
Another consideration concerns the fiber-optic NLSE from which our study originates. 
Indeed we discussed a sort of ``basic'' model \cite{A}, but we could have considered an equation, usually employed to describe transmission in multi-span fiber-optic links, with piecewise constant, or periodic, dispersion parameter $\beta_2$.
In this case Painlev\'e analysis is still applicable and one should investigate how the function $V_2$, following from relation \eqref{relation2} looks like, and how the error estimates gets modified. 
This point is certainly worth to be addressed in a future work.
\par
To prove the result we had to prove that the function $t\mapsto t^2v(z,t)$ has finite integral $\int_{\RR}|t^2v(z,t)|^2 dt$ 
for every $z$ in a suitable interval of $\RR$, where $v(z,t)$ is a solution of Eq.\ \eqref{modellointegrabile}. If we have chosen functions different to zero for the arbitrary $V_0(z)$ and $V_1(z)$, the result also holds true, but the error bound should be grater than that stated in the theorem, because of additional terms in the relation \eqref{duhamel}. 
To get $\int_{\RR}|t^2v(z,t)|^2 dt<\infty$, we used the transformation that allows to pass from Eq.\ \eqref{modellointegrabile} to 
Eq.\ \eqref{modellocubica} \cite{He}, so that we were allowed to work with the standard cubic NLSE. 
In Ref.\ \cite{C} is proven that the function $T\rightarrow T Q(Z,T),$ where $Q(T,Z)$ is solution of the classical standard cubic 
NLSE \eqref{modellocubica}, belongs to $L^2$. 
However, this was not enough for our aim: in order to use Lemma \ref{shift} we needed to prove the finiteness of the integral 
$\int_{\RR}\big(T^2Q(Z,T)\big)^2 dT,$ for every $Z$ in a suitable bounded interval of $\RR$. 
In the framework of the standard cubic NLSE, by some tools usually employed to prove finite-time blow up of solutions \cite{C}, 
we proved that the function $T\mapsto TQ_T(Z,T)$ belongs to $L^2(\RR)$ for every $Z$ in a suitable bounded interval of $\RR$.
Using this result we proved that the function $T\mapsto T^2Q(Z,T)$ belongs to $L^2(\RR)$ for every $Z$ in a suitable bounded 
interval of $\RR$.
\par 
In the proofs we carried out estimates for $||TQ_T(Z,T)||_{L^2}$ and $||T^2Q(Z,T)||_{L^2}$, uniformly with respect to $Z$ 
in an interval.
For this reason, when returning to Eq.\ \eqref{modellointegrabile} we obtain Theorem \ref{maintheorem} for $z$ in a suitable interval.
We remark that the result is independent of the sign of the character of the NLSE: focusing (if $\rho=1$) or defocusing ($\rho=-1$).
\par
The standard cubic NLSE, in the focusing case, possesses special analytic solutions: the soliton-like ones. 
So we can obtain analytic solitonic solutions of Eq.\ \eqref{modellointegrabile} by means of above mentioned 
transformation \cite{He}. 
Theorem \ref{maintheorem} provides an estimate of the distance between particular solutions of Eq.\ \eqref{modello} 
and particular solutions of Eq.\ \eqref{modellointegrabile}, depending on $z$ and on the initial datum. 
Therefore, we can obtain solutions of Eq.\ \eqref{modello} that are close to soliton-like solutions of Eq.\ \eqref{modellointegrabile}
in the $L^2$-norm. 
It is well-known that solitonic solutions of standard NLSE \eqref{modellocubica} are stable by considering a particular distance 
that involves $H^1$-norm \cite{Z}. 
Thanks to the transformation \eqref{trasformazionimodello}, this kind of stability can be passed to  Eq.\ \eqref{modellointegrabile}
and, then, we have straightforwardly both a definition of stability and conditions on the solitonic solutions of 
Eq.\ \eqref{modellointegrabile} to be stable \cite{F}. 
\par
Theorem \ref{maintheorem} could be improved by considering the $H^1$-norm; in this way, it is possible to construct a new 
distance that would allow to establish conditions under which the solutions of Eq.\ \eqref{modello} that are close to solitonic 
solutions of  Eq.\ \eqref{modellointegrabile} are stable, that is, close (in a suitable sense) to the initial datum during the evolution. 
Unfortunately, because of the dependence of the closeness estimations on the evolution variable $z$, we expect that this kind 
of stability can be proved only for small $z$. 
This will constitute the object of study of a forthcoming investigation.

\section*{Acknowledgments}
This work was completed during D.F.'s stay at Universit\`a di Firenze with a PhD grant from CNIT Laboratories 
(Consorzio Nazionale Interuniversitario per le Telecomunicazioni) of Scuola Superiore Sant'Anna (Pisa).
We also wish to thank Riccardo Adami (Politecnico di Torino) for his bright remarks and especially for his encouragement.

\end{document}